# Positioning for Visible Light Communication System Exploiting Multipath Reflections

Hamid Hosseinianfar*, Mohammad Noshad† and Maite Brandt-Pearce‡
*‡Charles L. Brown Department of Electrical and Computer Engineering,
University of Virginia, Charlottesville, VA 22904.
†VLNCOMM LLC, Charlottesville, VA 22904
Email: *hh9af@virginia.edu, †noshad@vlncomm.com, ‡mb-p@virginia.edu

*Abstract*—In this paper, we introduce a new uplink visible light indoor positioning system that estimates the position of the users in the network-side of a visible light communications (VLC) system. This technique takes advantage of the diffuse components of the uplink channel impulse response for positioning, which has been considered as a destructive noise in existing visible light communication positioning literature. Exploiting the line of sight (LOS) component, the most significant diffusive component of the channel (the second power peak (SPP)), and the delay time between LOS and SPP, we present a proof of concept analysis for positioning using fixed reference points, i.e. uplink photodetectors (PDs). Simulation results show the root mean square (RMS) positioning accuracy of 25 cm and 5 cm for one and 4 PDs scenarios, respectively.

## I. INTRODUCTION

Ubiquitous use of light-emitting diodes (LEDs) alongside their fast modulation capability paves the way for introducing new indoor services such as internet of things, visible light communications (VLC) and positioning through lighting. In comparison with radio-frequency (RF) indoor networks, visible light communication systems are considered as a viable approach for indoor wireless data access providing higher security and larger multiuser capacity [1]. Indoor positioning is difficult to achieve because the signals from GPS satellites can be blocked in indoor areas and underground facilities. VLC systems can provide centimeter accuracy in finding the location [2]. Various positioning algorithms have been proposed using visible light signals that demonstrate higher accuracy compared to conventional RF techniques [3], [4].

The main approaches for estimating the users' location in visible light positioning systems generally rely on three features of the received signals: time of arrival (TOA), angle of arrival (AOA), and received signal strength (RSS). RSS-based techniques use the intensity of the signal to estimate the distance from the transmitter to the receiver. These techniques have the potential to achieve a high accuracy in visible light positioning systems because of the strong line of sight (LOS) signals, which are often not available in RF systems. However, the accuracy of the RRS techniques that rely on LOS signals is limited due to the shadowing and multi-path effects, which make the relationship between the distance and RSS unpredictable [2], [5], [6]. In TOA-based systems, the position of the receiver is estimated by measuring the arrival time of the signal from different transmitters, and hence, it requires perfect synchronization between the transmitters, which can limit the application of these systems. Theoretical limits have been presented on the accuracy of the TOA-based [4] and RSS-based [7] positioning techniques. TOA and RSS locate users using triangulation, trilateration, and/or fingerprinting and in general, require at least two LEDs when the height of the user is known, and three LEDs when it is unknown. AOA has been used with imaging receivers to locate the users by measuring the angle at which the line-of-sight (LOS) signal from the transmitter is received. Shadowing can interrupt the performance of this type of positioning system. AOA-based techniques can estimate the user's location using one imaging receiver when the height of the user is known and two imaging receivers when it is not [2].

In this paper, we propose a new positioning technique for indoor visible light systems that uses the characteristics of the channel impulse response to locate users. This method exploits the multipath signals to get a better estimate of the user's position rather than considering them as noise. While most of the research on visible light positioning has concentrated on downlink and user-side positioning, the proposed algorithm is developed for the infrared (IR) uplink of a VLC system. This localization algorithm provides the network the information needed for communications purposes, such as resource allocation and hand-off where the network monitors the number and position of the users. For this purpose, less complex algorithms are preferred over more accurate ones. In the proposed algorithm, one reference point, i.e., one uplink receiver, suffices for estimating the user's position. However, adding more reference points enhances the positioning accuracy.

The rest of the paper is organized as follows. In Section II, the system model and measurement scenario are discussed. The positioning algorithm is presented in Section III. The numerical results are presented and discussed in Section IV. Finally, the paper is concluded in Section V.

## II. SYSTEM DESCRIPTION

### A. Fingerprint Definition

In a typical visible light communication system, there are usually multiple white LED fixtures on the ceiling that transmits downlink data and multiple infrared photo-detectors (PDs) that capture uplink signals as shown in Fig. 1. In this paper we propose a technique to find the position of the users using only one transmitter-receiver pair. We further, extend this algorithm to multiple receivers for the sake of accuracy. The idea in this work is to use the channel impulse response of the uplink channel to locate the user, and map the characteristics of the impulse response to the location of the user. Fig. 2 illustrates the room configuration considered for the proposed uplink positioning system. In this model, a PD is assumed to be installed on the ceiling at the point $(x^{(r)}, y^{(r)}, h^{(r)})$ facing vertically downwards, and the user at coordinates $(x, y, h)$ is assumed to have an infrared LED transmitter that is facing vertically upwards.

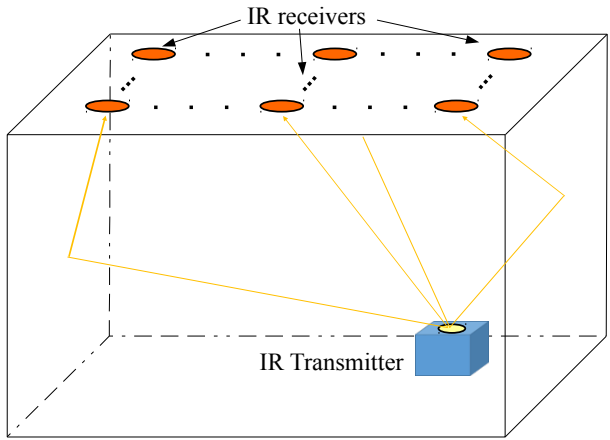

Fig. 1. System configuration for visible light communication uplink system

To develop the proposed positioning algorithm, we first divide the indoor area into a grid and then create a database of the channel impulse responses for different positions of the user on the grid, $C_k = (x_k, y_k), k \in \{1, 2, \ldots, MN\}$, for a known height $h$. We consider an $N \times M$ grid, as shown in Fig. 2. We focus on the main features of the impulse response, which give sufficient information about the user location. For the proof of concept, we consider the LOS peak power $P_{LOS}$, the second power peak (SPP) term $P_{SPP}$, i.e., the first peak of the diffuse term, and the arrival time difference between these two components $\Delta \tau$. The vector $S_k = [P_{LOS}, P_{SPP}, \Delta \tau], k \in \{1, 2, \ldots, MN\}$, represents the constellation vector corresponding to the $k$th point on the measurement grid for one PD. For the multiple PDs scenario, the feature space is simply expanded to $Q \times 3$ dimensions, where $Q$ is the number of PDs.

*B. Channel Model*

In indoor environments, the channel response of optical wireless systems includes both LOS and diffuse components that are caused by multipath reflections. For simplicity, in our simulation we consider only the first bounce of the channel's diffuse components. This approximation has a small error since the first bounce is the strongest component of the diffuse part. In addition, considering that the paths with more bounces have larger delays, neglecting the paths with more than one reflection does not have significant impact on the second power peak [8]. The impulse response is captured in simulation through ray tracing, presented in [9]. Fig. 3 illustrates the impulse response for three different user locations. For a fixed IR receiver position on the ceiling, each point of the room has a unique channel response and the LOS peak power, SPP, and their time delay can uniquely characterize the user's position. Table I presents the simulation parameters considered for this work.

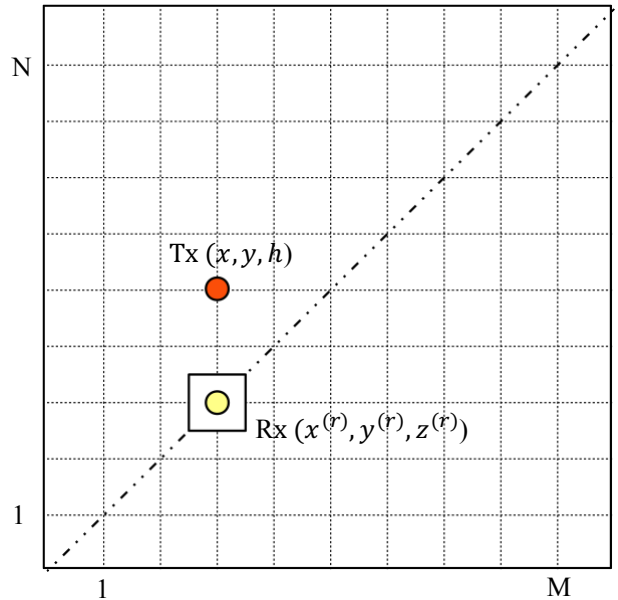

Fig. 2. Birds-eye view of the room.

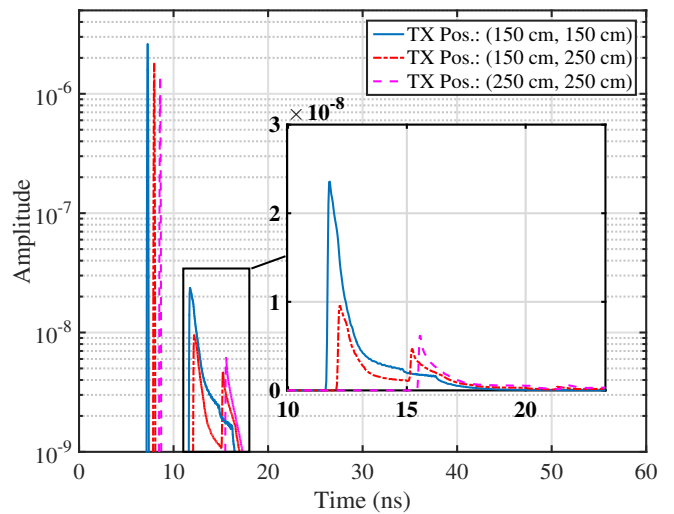

Fig. 3. Impulse response of the channel for different locations of the transmitter when the receiver is located at $(150 \text{ cm}, 150 \text{ cm}, 300 \text{ cm})$ and the simulation parameters are as in Table I.

TABLE I
SIMULATION PARAMETERS OF RAY TRACING CHANNEL MODEL [9].

| Transmitter Parameters | Value |
|---|---|
| Height | 0.85 m |
| Wavelength | 950 nm |
| Lambertian Mode (m) | 1 |
| LED transmit power, $P_T$ | 10 mW |
| **Receiver Parameters** | **Value** |
| Surface area of the PD, $A_{PD}$ | 1 cm$^2$ |
| Height | 3 m |
| Field of View (Half Angle) | 70° |
| Wall reflectance coefficient, $\rho$ | 0.8 |
| Reflecting element area, $A_{ref}$ | 2 cm |
| **Room Parameters** | **Value** |
| Room Size | $5 \times 5 \times 3$ m$^3$ |

## III. POSITIONING ALGORITHM

In this section, we describe our indoor positioning algorithm. Given that the transmitter sends an ideal delta function $\delta(t)$, the received signal is the channel impulse response with additive white Gaussian noise (AWGN) that can be defined as

$$v(\boldsymbol{\theta}, t) = h(\boldsymbol{\theta}, t) + n(t) \tag{1}$$

where $h(\boldsymbol{\theta}, t)$ is the corresponding channel impulse response and $n(t)$ is AWGN. In this algorithm, the user's position is estimated using features captured from $v(\boldsymbol{\theta}, t)$: the received peak power from LOS, and SPP components of the impulse response, and the time delay between LOS and SPP components. These three observations can be expressed as

$$\begin{aligned} v_1(\boldsymbol{\theta}) &= P_{\text{LOS}}(\boldsymbol{\theta}) + n_1, \\ v_2(\boldsymbol{\theta}) &= P_{\text{SPP}}(\boldsymbol{\theta}) + n_2, \\ v_3(\boldsymbol{\theta}) &= \Delta\tau(\boldsymbol{\theta}) + n_3. \end{aligned} \tag{2}$$

where $\boldsymbol{\theta} = (x, y)$ is the two dimensional coordinates of the user to be estimated, and $n_1$ and $n_2$ are independent Gaussian noises with variance $\sigma^2$. $n_3$ is the time delay noise between LOS and SPP which is modeled as a Gaussian noise with variance $\sigma_\tau^2$ and assumed independent from $n_1$ and $n_2$. Let $\boldsymbol{V}(\boldsymbol{\theta}) = [v_1(\boldsymbol{\theta}), v_2(\boldsymbol{\theta}), v_3(\boldsymbol{\theta})]^T$ be the observation vector.

### A. Calculation of Time Jitter Noise $\sigma_\tau^2$ for Peak Detector

In real applications, the observation components defined in (2) are the output of a peak detector. The variance $\sigma^2$ of peak amplitude noises, namely, $n_1$ and $n_2$, are the same as the received signal variance. However, the variance of time delay noise $n_3$, the time jitter noise of a peak detector, is a function of $\sigma^2$. To figure out the relation between the time jitter and amplitude noise, we can simplify the problem to the relation of amplitude noise and time noise of the zero crossing, when the amplitude noise distribution is Gaussian. The first derivative of the received signal in (1), $v'(t) \doteq \frac{\partial v(\boldsymbol{\theta}, t)}{\partial t}$ crosses zero at the times near the LOS and SPP peaks, with the same amplitude noise variance. At these zero crossing times, the function $v'(t)$ can be approximated as linear with constant slope $\beta = \mathrm{d}v'(t)/\mathrm{d}t|_{v'(t)=0}$. Thus, the amplitude noise can be translated to jitter noise $n_3$. This linear approximation is accurate for high SNR scenarios. Considering this approximation, the distribution of the jitter noise remains Gaussian. The corresponding noise variance can be written

$$\sigma_\tau^2 = \frac{\sigma^2}{\beta^2}. \tag{3}$$

In this paper, we use $\beta^2 = 30$, based on our numerical results.

### B. Maximum Likelihood Estimation

Maximum likelihood estimation is employed for positioning, where the algorithm looks for the constellation point $S_k(\hat{\boldsymbol{\theta}}) = [P_{\text{LOS}}(\hat{\boldsymbol{\theta}}), P_{\text{SPP}}(\hat{\boldsymbol{\theta}}), \Delta\tau(\hat{\boldsymbol{\theta}})]$, $k \in \{1, \ldots, MN\}$ that has the minimum Euclidean distance to the observation vector $\boldsymbol{V}(\boldsymbol{\theta}) \in$ observation plane. $\hat{\boldsymbol{\theta}} \in$ room plane corresponding to this constellation point is considered as the estimated location. Fig. 4 illustrates an example of mapping between some region close to $\boldsymbol{V}(\boldsymbol{\theta})$ in the observation plane onto the room plane. The decision regions are depicted based on the minimum Euclidean distance criterion. In this example, $\hat{\boldsymbol{\theta}} = C_i$ is the estimated location since $\boldsymbol{V}(\boldsymbol{\theta})$ is located in the $S_i$ decision region (see Fig. 4).

### C. Calculation of the Positioning Error

Considering the minimum distance algorithm, the probability of choosing $S_i(\hat{\boldsymbol{\theta}})$ given the position of $\boldsymbol{\theta}$ is denoted by

$$\epsilon_j = Pr\{\boldsymbol{S}_{j=\hat{i}}(\hat{\boldsymbol{\theta}})|\boldsymbol{\theta}\}, \tag{4}$$

where

$$\hat{i} = \arg\min_i \left|\boldsymbol{V}(\boldsymbol{\theta}) - \boldsymbol{S}_i(\hat{\boldsymbol{\theta}})\right|^2 \quad i \in \{1, 2, \ldots, NM\} \tag{5}$$

is the index of the selected center point, $S_i(\hat{\theta})$. The root mean square (RMS) positioning error can be calculated as

$$d_{RMS} = \sqrt{\mathbf{E}_\theta\left\{\sum_{j=1}^{N \times M} |\boldsymbol{\theta} - \boldsymbol{C}_j|^2 \epsilon_j\right\}} \tag{6}$$

where $\mathbf{E}_\theta\{\cdot\}$ denotes the expectation over different locations $\boldsymbol{\theta}$.

Due to the complex shapes of the decision regions, a derivation of the exact value of $\epsilon_j$ is not trivial. In this case, a reasonable approach, especially for high SNR scenarios, is to calculate a lower-bound (LB) on the estimation error [10]. To that end, we consider only $\epsilon_j$'s that correspond to the two closest center points to the observation point, $\boldsymbol{V}(\boldsymbol{\theta})$. Then, (4) evaluated at these center points can be rewritten

$$\begin{aligned} \epsilon_i &= Q(\sqrt{\tfrac{\mathbf{L}^T(i) \cdot \boldsymbol{\Sigma}^{-1} \cdot \mathbf{L}(i)}{2}}), \\ \epsilon_{i'} &= 1 - Q(\sqrt{\tfrac{\mathbf{L}^T(i) \cdot \boldsymbol{\Sigma}^{-1} \cdot \mathbf{L}(i)}{2}}). \end{aligned} \tag{7}$$

where $\mathbf{L}(i)$ is the distance vector between $\boldsymbol{V}(\boldsymbol{\theta})$ and the boundary between the two closest center points, named $i$ and $i'$ (see Fig. 4). The covariance matrix can be defined as

$$\boldsymbol{\Sigma} = diag(\sigma^2, \sigma^2, \sigma_\tau^2). \tag{8}$$

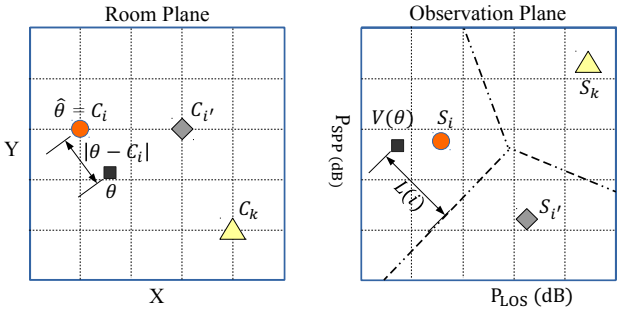

Fig. 4. Mapping the room plane onto the observation plane. The $S_i$, $S_{i'}$, and $S_k$ are three typical observation points, close together in observation space and, the $C_i$, $C_{i'}$, and $C_k$ are corresponding center points in the room plane.

The $\sigma^2$ and $\sigma_\tau^2$ are the variances of the noises on the observation vector defined in (2). For the multiple PDs scenario, the observation space vectors are simply expanded to $Q \times 3$ dimensions by concatenating the corresponding vectors from different PDs. In this case, $\Sigma$ is the $3Q \times 3Q$ covariance matrix formed by values of $\sigma^2$ and $\sigma_\tau^2$ in the appropriate places on the diagonal.

## IV. NUMERICAL RESULTS

In order to model the multipath signals in a typical room, simulations are done in an empty cubicle of size $5 \times 5 \times 3$ $m^3$. Table I summarizes the parameters used in the simulations. In these simulations, we assume the single PD to be located at point $(1.5, 1.5, 3)$ on the ceiling. For multiple PDs scenarios, the other PDs are located at points $(3.5, 1.5, 3)$, $(1.5, 3.5, 3)$, and $(3.5, 3.5, 3)$. The transmitter is assumed to be at 85 cm from the floor.

Fig. 5 demonstrates the contour plots for the three observation components, namely, $P_{\text{LOS}}$, $P_{SPP}$ and $\Delta\tau$. Based on previously presented positioning methods, like trilateration, given enough power measurements (three in trilateration), the user position can be estimated to be at the intersection point of power observation contour plots. As an intuition for the proposed algorithm, here the user position can be estimated as the intersection of three observation contours $P_{\text{LOS}}$, $P_{SPP}$, and $\Delta\tau$.

Fig. 6-(a), and (b) illustrate the constellation points in the room plane and in the corresponding observation plane, based on two observations components, $P_{\text{LOS}}$ and $P_{SPP}$. The corresponding points in both planes are denoted with markers of the same size and color. As shown in Fig. 6-(b), the mapping to the observation plane maintains the same room plane pattern, i.e., almost all the point on the observation plane are surrounded by the mapping of its corresponding neighbors in the room plane. Hence, maximum likelihood estimation, i.e., the minimum Euclidean distance in the observation plane, leads to the closest measurement point on the room plane. Adding the third observation component, $\Delta\tau$, we can see in

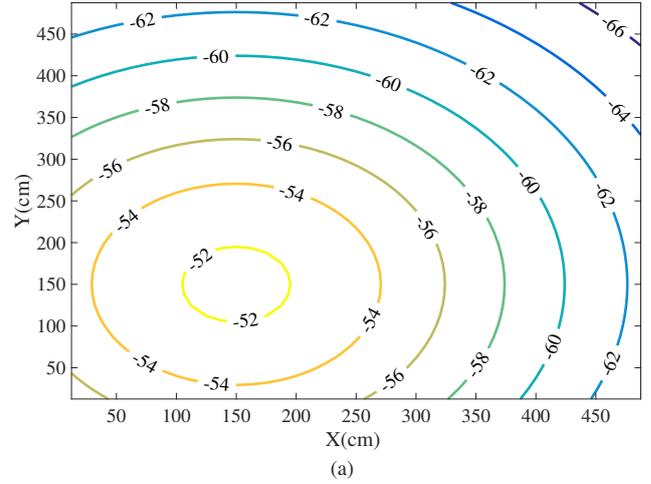

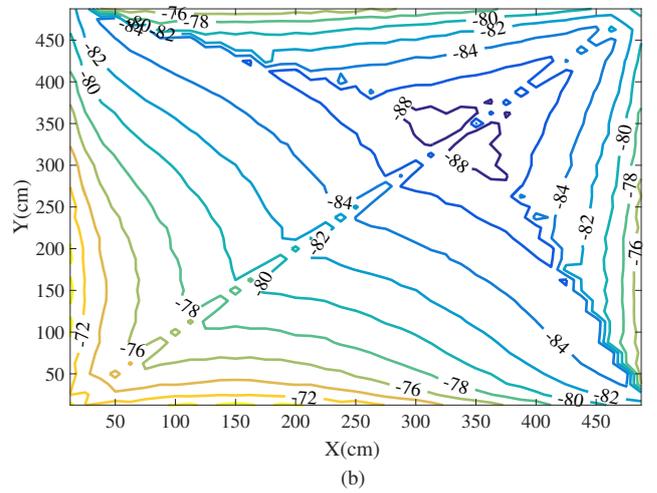

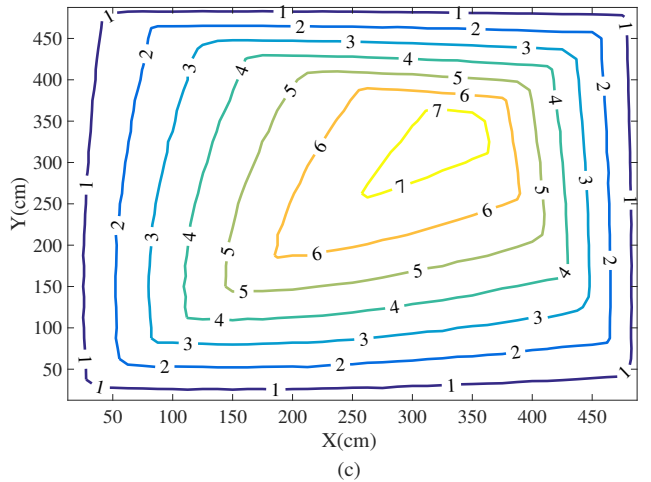

Fig. 5. Contour plots in room space: a) LOS component of received power (dB), b) SPP component of received power (dB), c) Time delay between the LOS and SPP components (ns).

Fig. 6-(c) that the distances between the points are increased, which brings higher positioning accuracy.

In order to determine the estimation error, we run a Monte Carlo (MC) simulation by randomly choosing the location of the user and estimating the closest grid point on the room plane using the proposed algorithm. The RMS positioning error is a representation of the probability of choosing a constellation point by the algorithm and the mapping cost of that decision. However, given that the algorithm tries to find the closet grid point, the larger the grid step size, the worse the absolute positioning accuracy becomes, and yet the lower the probability of mapping to some far bin in the room. Therefore, there is a trade off between the room plane absolute grid accuracy and the cost of wrong mapping. Figs. 7-(a) and 7-(b) demonstrate the RMS positioning errors for the 2 and 3 observations algorithms, respectively. For a different number of PDs, the MC simulations converge to the LB results (calculated in section III-C) at high SNR, which verifies the simulation results. Deploying more PDs leads to higher accuracy level: for 4 PDs we reach the minimum error that is inevitable due to quantization on a grid. As expected, for the same SNR and number of PDs, the 3 observations algorithm provides better accuracy compared to the 2 observations algorithm.

Fig. 8 shows the RMS positioning error versus grid step for a high SNR scenario. There is a nearly linear relation between accuracy and the grid step size for all multiple PDs scenarios. In addition, for a larger number of PDs, the RMS error gets closer to the RMS quantization error, where the estimated location is mapped to the closest grid point in the room, in the high SNR case.

*A. Addressing Practical Challenges*

The performance of the proposed algorithm tightly depends on the channel model. Therefore, the features extracted from the impulse response are different for real indoor spaces containing furnishings, surfaces with different reflection factors, specular reflections from windows, etc. In real scenarios, it is only a matter of defining the fingerprint database: the actual impulse response can be learned once for a specific grid, and then the algorithm can be deployed. The other practical concern has to do with the shadowing effect, which is inevitable when there are moving users and/or objects in the room. In this case, installing multiple PDs on the ceiling can address the problem. Tracking the movement path of the users can increase the accuracy of the proposed algorithm by limiting the number of constellation points for the user's location. The positioning error can also be reduced by redesigning the grid in the room plane and optimizing the observation constellation using irregular points in the room, and more regular constellations in the observation space.

## V. CONCLUSION

In this paper, an infrared uplink positioning algorithm is proposed that takes advantage of the user location information embedded in the multipath reflections. Using the LOS, SPP and the time delay between these two components, a database of room fingerprints on a grid is created. The performance of the maximum likelihood estimation algorithm is evaluated for several different grid steps and multiple PDs scenarios. The results show a RMS positioning error of 5 cm using 4 PDs, a SNR of 50 dB, and grid step of 14 cm.

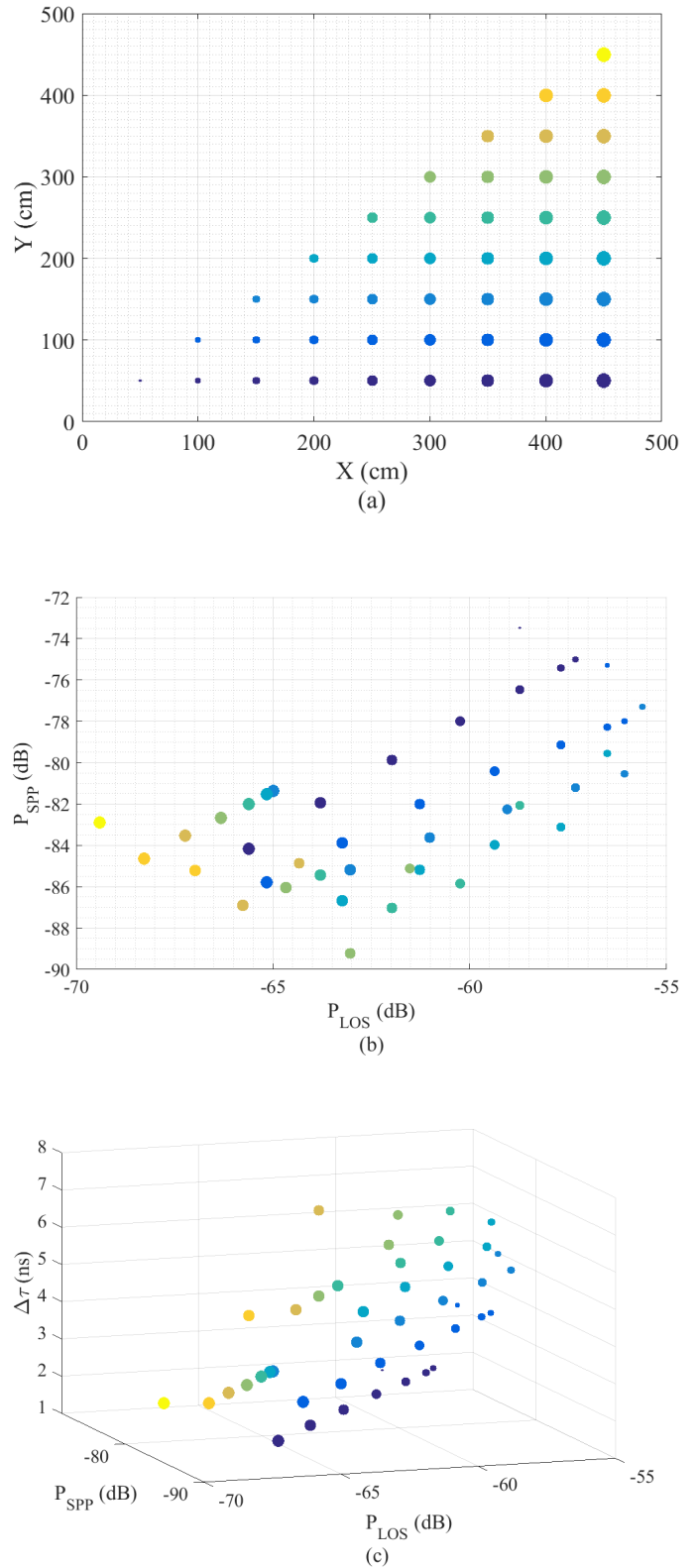

Fig. 6. a) Anchor points in room plane, b) Corresponding fingerprints in observation plane for two observations, $P_{\text{LOS}}$ and $P_{SPP}$, c) Corresponding fingerprints in observation plane for three observations, $P_{\text{LOS}}$, $P_{SPP}$, and $\Delta\tau$.

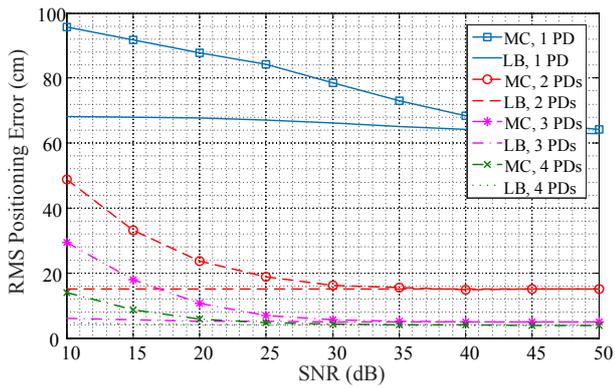

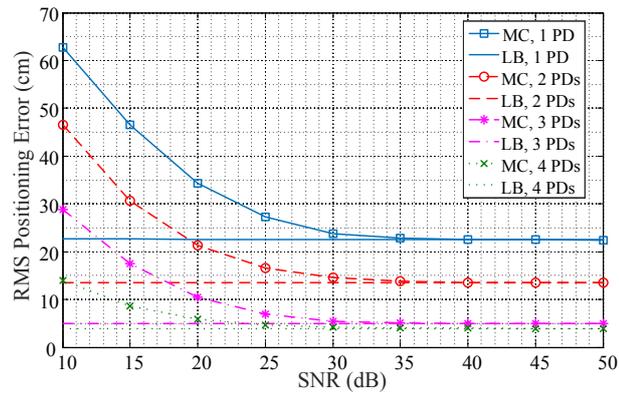

Fig. 7. RMS distance error for a grid step of 14 cm and multiple PDs scenarios: a) Two observations, b) Three observations.

## VI. Acknowledgement

This work was funded by the National Science Foundation (NSF) through the STTR program, under award number 1521387.

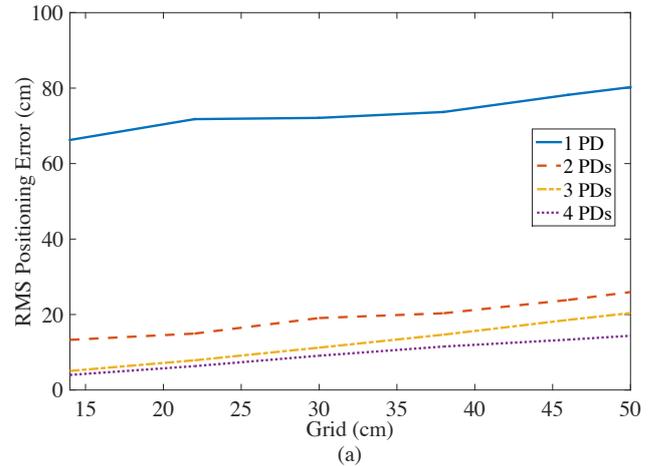

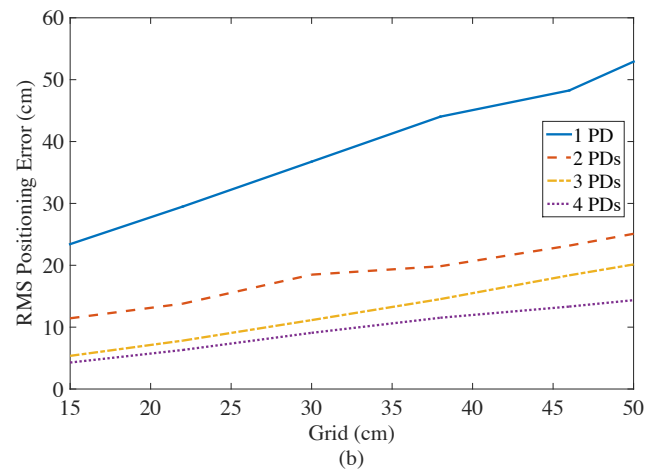

Fig. 8. RMS distance error for SNR= 50 dB and multiple PDs scenarios: a) Two observations, b) Three observations.